\documentclass[12pt]{iopart}

\usepackage{graphicx}
\usepackage{color}

\def\Journal#1#2#3#4{{#1} {\it #2}, {\bf #3} {#4}}
\def\PRL{ Phys. Rev. Lett.}
\def\PLB{ Phys. Lett. B}

\begin{document}

\title[Charm cross-section measurements at STAR]
{Scaling of the charm cross-section and modification of charm $p_{T}$ spectra at RHIC}

\author{Chen Zhong for the STAR Collaboration}

\address{Shanghai Institute of Applied Physics, CAS, Shanghai 201800, P. R. China}
\ead{zhongchen@sinap.ac.cn}

\begin{abstract}
\label{abstra}

Charm production from the direct reconstruction of $D^0$ ($D^0\to
K\pi$ up to 2 GeV/$c$) and indirect lepton measurements via charm
semileptonic decays ($c\to e+X$ at 0.9\textless$p_T$\textless5.0
GeV/$c$ and $c\to \mu+X$ at 0.17\textless$p_{T}$\textless0.25
GeV/$c$) at $\sqrt{s_{_{NN}}}=200$ GeV Au+Au collisions are
analyzed. The transverse momentum ($p_T$) spectra and the nuclear
modification factors for $D^0$ and for leptons from heavy flavor
decays is presented. Scaling of charm cross-section with number of
binary collisions at $\sqrt{s_{_{NN}}}=200$ GeV from d+Au to Au+Au
collisions is reported.
\end{abstract}

\section{Introduction}
\label{intro}
In relativistic heavy-ion collisions, charm quarks are believed to
be produced at early stages via initial gluon fusion and their
production cross-section can be evaluated by pQCD~\cite{cacciari}.
Study of the binary collision ($N_{bin}$)
scaling properties of the charm total cross-section in p+p,
d+Au to Au+Au collisions can test if heavy-flavor quarks are
produced exclusively at initial impact~\cite{ffcharm}. Due to the
heavy mass of charm quarks, charmed hadrons might freeze out
earlier than light flavor hadrons.
Charm energy loss, highly sensitive to the properties of medium,
can be inferred by studying the nuclear modification factor of its semileptonic
decayed electron.

\section{Experiment and Analysis}
\label{exp}
The data used for this analysis were taken with the Time Projection Chamber (TPC)
and the Time Of Flight (TOF) detectors in the STAR~\cite{STAR_NIM} experiment
during the $\sqrt{s_{_{NN}}}=200$ GeV Au+Au run in 2004. The TPC is the main
tracking device in STAR, which provides particle identification within a
pseudorapidity coverage of $|\eta|\textless$1.5 and full azimuthal coverage~\cite{TPC_NIM}.
In this study the measurements of the ionization energy loss (dE/dx) of charged tracks in
the TPC gas is used to identify pions, kaons, electrons and muons. The TOF, which measures
the velocity of charges particles, covers $\pi$/30 rad in azimuth and -1$\textless\eta\textless$0 in pseudorapidity
at a radius of $\sim$ 220 cm from the beam pipe~\cite{TOF_PLB}. About 7.8 million
Au+Au events with 0-80\% centrality and 15 million top 12\% central Au+Au collision
events were used in the analysis.
\begin{figure}
\centering
  \includegraphics[scale=0.82]{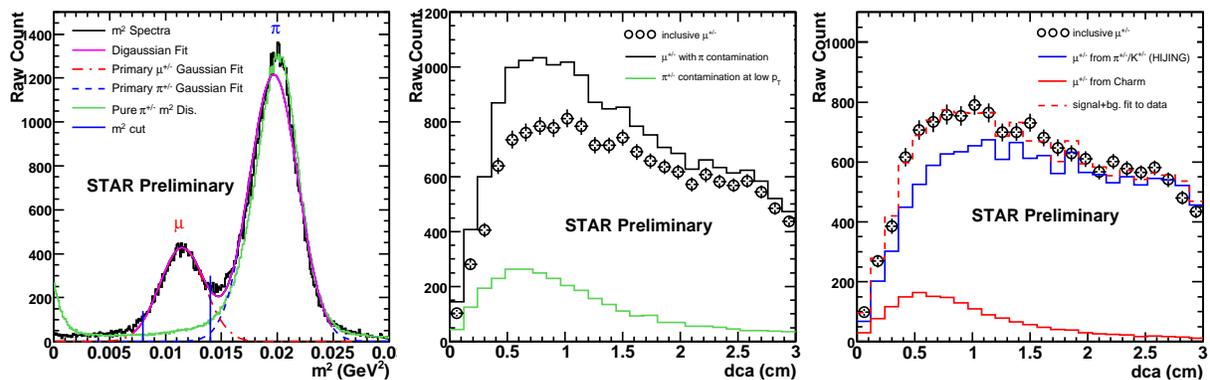}
\caption{Left panel: Particle mass squared
distribution ($m^{2} = (p/(\beta\gamma))^{2}$). The muon and pion peaks are clearly visible.
Middle panel: Procedure to remove the residual pion background.
Right panel: DCA distributions of muons from charm decays and from weak $\pi/K$ decays (from HIJING simulation)}
\label{fig:1}
\end{figure}
For the hadronic decay mode, reconstruction of $D^{0}\to K^{-}\pi^{+}(\bar{D}^{0}\to {K}^{+}{\pi}^{-})$
(branching ratio of 3.8\%) was carried out.
An alternative method to study charm production is through the measurement of
from semileptonic electrons/muons decays of charmed hadrons ($c\to e/\mu+X$ with a branching ratio of 6.87\%/6.5\%).
~\cite{ffcharm,SQM06Pro,PDG}.
Lepton identification was carried out using the STAR TPC in conjunction with TOF.
The single muon measurements benefit from the absence of Dalitz decays and photon conversions present in the
electron channel. We have carried out muon measurements in the $p_{T}$ region 0.17\textless$p_{T}$\textless0.25 GeV/$c$
in both 0-80\% and top 12\% central Au+Au collisions at $\sqrt{s_{_{NN}}}=200$ GeV.

The left panel of Fig. \ref{fig:1} shows the $m^{2}=(p/(\beta\gamma))^{2}$
distribution from TOF after TPC $dE/dx$ selections.
A clear muon peak is observed within a mass window of $0.008\textless m^{2}\textless 0.014$.
We also see some residual pion background in the mass range. The residual pions are removed
statistically by studying the distance of closest approach (DCA) of the tracks from
the collision vertex within the above mass range. The method and the resultant inclusive
muon DCA distribution (open circles) are shown in the middle panel of Fig. \ref{fig:1}.
The right panel illustrates the procedure to obtain muon yields from charm
semileptonic decays from inclusive muon DCA distribution. This is done statistically by removing
the contribution of muons from $\pi/K$ weak decays. We obtain the $\pi/K\to\mu$ DCA distributions
from HIJING~\cite{HIJING} simulations using the full STAR Detector configuration.
We then use DCA of muons from primary particles and those coming from weak decays
of $\pi/K$ (HIJING simulation) to fit the inclusive muons DCA spectra.
This is used to get the raw yields of muons from charm semileptonic decays.

\begin{figure}
\centering
  \includegraphics[scale=0.30]{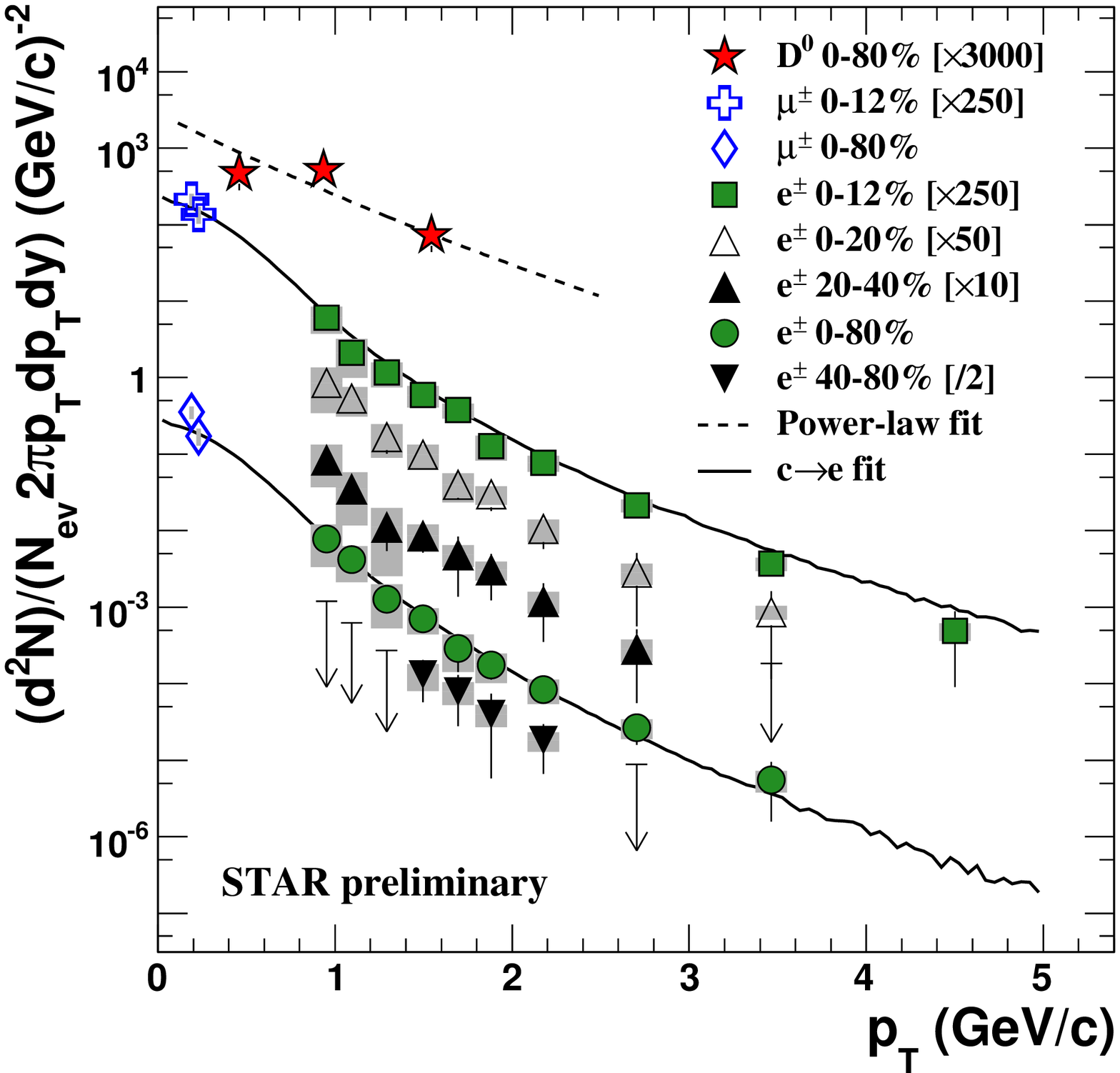}  \includegraphics[scale=0.30]{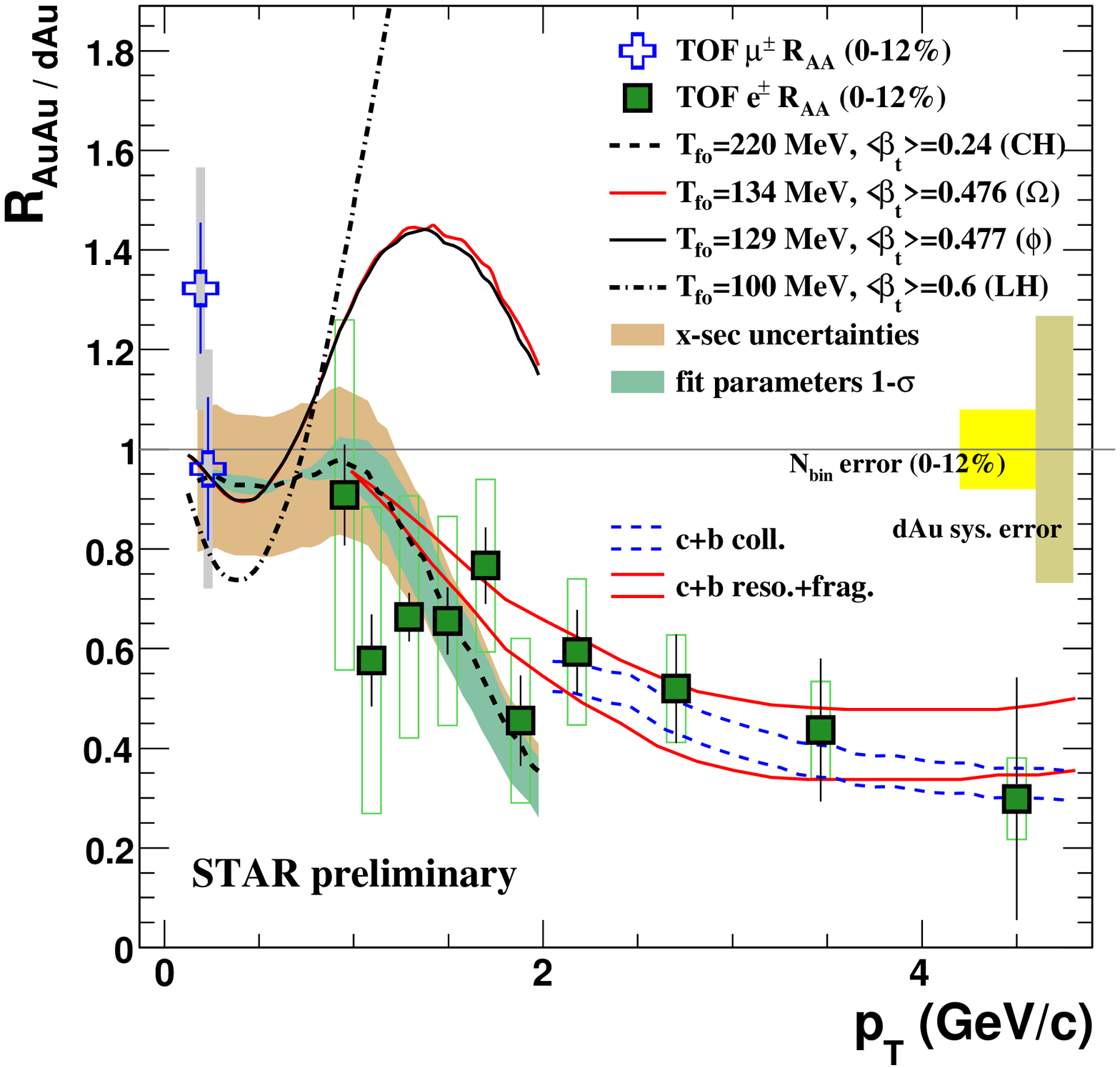}
\caption{Left Panel: The $p_{T}$ spectra of $D^{0}$ and electron/muon from charm semileptonic
decays in Au+Au collisions.
Dashed and solid curves are power-law combined fit for $D^{0}$ and decayed leptons, respectively.
Right Panel: Nuclear modification factor ($R_{AuAu/dAu}$) of electron and muon as
a function of $p_{T}$. Low $p_{T}$ muon $R_{AuAu/dAu}$ is consistent with number of
binary scaling and intermediate $p_{T}$ electron $R_{AuAu/dAu}$ show a strong suppression.}
\label{fig:2}
\end{figure}

\section{Results}
\label{analy}
The left panel of Fig. \ref{fig:2} shows invariant yields for $D^{0}$ (stars) and electrons/muons from
charm semileptonic decays as a function of $p_{T}$.
A power-law function was used to fit the $D^{0}$ spectrum combined with the lepton
spectra from charmed hadron semileptonic decays.
All three measurements together stringently constrain the charm cross-section at RHIC.
The right panel of Fig. \ref{fig:2} shows the nuclear modification
factors ($R_{AuAu/dAu}$) as a function of $p_{T}$ for various collision centralities which can give insight
into the particle production mechanism.
The $R_{AuAu/dAu}$ for non-photonic electron and muon production are derived by using the
$N_{bin}$ scaled $p_{T}$ spectra in central Au+Au collisions divided by the $N_{bin}$ scaled decayed
electron spectra from a combined fit in d+Au collisions~\cite{SQM06Pro,dAuPRL}.
Those are shown as open squares and crosses, respectively.
Considering the extrapolation of the d+Au fit to lower momenta, muons seen to follow a $N_{bin}$ collision
scaling within the systematical uncertainties.
The non-photonic electron $R_{AuAu/dAu}$ is suppressed as strongly as that of light hadrons~\cite{PRL97152301}, which indicates
that charmed hadrons experience energy loss in the medium.
Model calculations~\cite{Ivan,Rapp} with different mechanism considering in-medium charm
resonances or charm diffusion and collisional dissociation of heavy mesons respectively can
reasonably describe the $R_{AuAu/dAu}$ for non-photonic electrons.
A blast-wave parameterization that assumes early kinetic
freeze-out of charmed hadrons (dashed curve) describes the $R_{AuAu/dAu}$ distribution better than those with the late
freeze-out assumption (black/red, dotted curves).

\begin{figure}
\centering
  \includegraphics[scale=0.30]{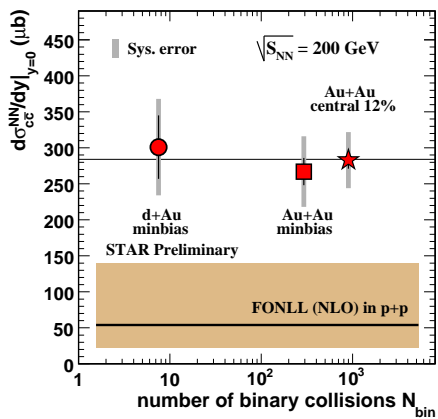}
\caption{Charm cross-section at mid-rapidity as a
function of number of binary collisions ($N_{bin}$) in d+Au,
0-80\% and 0-12\% central Au+Au collisions.}
\label{fig:Xsection}
\centering
\end{figure}

In Fig. \ref{fig:Xsection}, the charm cross-section extracted from a combination
of the three measurements is shown as a function of $N_{bin}$.
It is 1.33$\pm$0.06(stat.)$\pm$0.18(sys.) mb in 0-12\%  and 1.26$\pm$0.09$\pm$0.23 mb in 0-80\% central Au+Au collisions at $\sqrt{s_{_{NN}}}$=200 GeV.
Within errors the charm cross-section is found to follow binary collisions scaling. This supports the
conjecture of charm quarks being produced at early stages in RHIC.
The prediction from a recent pQCD calculation (FONLL) for p+p collisions is depicted by the band in Fig. \ref{fig:Xsection}~\cite{cacciari}. It underestimates the observed cross-section by a factor of 5.

\section{Conclusions}
\label{concl}
\vspace{-0.3cm}
We have reported the first measurement of single muon yields from charm semileptonic decays at low $p_{T}$ in Au+Au
collisions at $\sqrt{s_{_{NN}}}=200$ GeV from STAR experiment. The $R_{AuAu/dAu}$ of the low $p_{T}$ muons show
$N_{bin}$ scaling and those for non-photonic electron show strong suppression at intermediate $p_{T}$. Charm
cross-sections are extracted from a combination of the three measurements covering $\sim90\%$ of the kinematic
range within the detector acceptance. The present measurements of the charm cross-sections in different
collision centralities for Au+Au collisions are significantly improved over the previous measurements from
non-photonic electrons and/or from directly reconstructed charmed hadron with low statistics. The charm
cross-section is found to follow number of binary collisions scaling, which is a signature of charm production at
the initial stage.

The author wishes to thank NSFC 10610285 and KJCX2-YW-A14 for
contributing to the local expenses.

\section*{References}

\end{document}